\documentclass[aps,pra,twocolumn,superscriptaddress,10pt]{revtex4-1}

\usepackage[matrix,frame,arrow]{xy}

\usepackage[titletoc,title]{appendix}
\usepackage{graphicx,epic,eepic,epsfig,amsmath,latexsym,amssymb,verbatim,color}
\usepackage{dsfont}
\usepackage{MnSymbol}
\usepackage{tikz}
\usepackage{graphicx}

\usepackage{url}
\usepackage{hyperref}
\hypersetup{colorlinks=true,citecolor=blue,linkcolor=blue,filecolor=blue,urlcolor=blue,breaklinks=true}

\usepackage{textcomp}

\usepackage{stmaryrd}
\usepackage{amsthm}
\usepackage{subcaption}
\usepackage{algorithm}
\usepackage[noend]{algpseudocode}

\newcommand{\bra}[1]{\ensuremath{\left\langle#1\right|}}
\newcommand{\ket}[1]{\ensuremath{\left|#1\right\rangle}}

\newcommand{\proj}[1]{\ket{#1}\!\bra{#1}}

\begin{document}


\title{Distinguishing Unitary Gates on the IBM Quantum Processor}


\author{Shusen Liu}
\email[]{Shusen88.liu@gmail.com}
\affiliation{School of Data and Computer Science, Sun Yat-sen University, Guangzhou, Guangdong, 510006, P. R. China}
\affiliation{Centre for Quantum Software and Information, Faculty of Engineering and Information Technology, University of Technology Sydney, NSW 2007, Australia}

\author{Yinan Li}
\email[]{Yinan.Li@cwi.nl}
\affiliation{Centre for Quantum Software and Information, Faculty of Engineering and Information Technology, University of Technology Sydney, NSW 2007, Australia}
\affiliation{Centrum Wiskunde \& Informatica and Research Center for Quantum Software, the Netherlands.}

\author{Runyao Duan}
\email[]{Runyao.Duan@uts.edu.au, duanrunyao@baidu.com}
\thanks{This work was mainly completed while the third author was in the University of Technology Sydney.}
\affiliation{Centre for Quantum Software and Information, Faculty of Engineering and Information Technology, University of Technology Sydney, NSW 2007, Australia}
\affiliation{Institute for Quantum Computing, Baidu Inc., Beijing 100193, China}


\date{\today}

\begin{abstract}
An unknown unitary gates, which is secretly chosen from several known ones, can always be distinguished perfectly. In this paper, we implement such a task on IBM's quantum processor. More precisely, we experimentally demonstrate the discrimination of two qubit unitary gates, the identity gate and the $\frac{2}{3}\pi$-phase shift gate, using two discrimination schemes -- the parallel scheme and the sequential scheme. We program these two schemes on the \emph{ibmqx4}, a $5$-qubit superconducting quantum processor  via IBM cloud, with the help of the $QSI$ modules [S. Liu et al.,~arXiv:1710.09500, 2017]. We report that both discrimination schemes achieve success probabilities at least 85\%. 
\end{abstract}


\maketitle

\section{Introduction}
The discrimination of quantum operations asks to identify an unknown quantum operation from a set of known ones. As a fundamental task in quantum information and computation, many interesting aspects have been discovered over the last two decades, see~\cite{acin2001statistical,PhysRevLett.96.200401,PhysRevA.72.052302,duan2007entanglement,DuanFengYing2009,PhysRevLett.100.020503,1751-8121-40-33-016,Watrous:2008:DQO:2017011.2017021,Chen:2010:ADQ:2011438.2011450} (and references therein) for a partial list. As applications, the discrimination of quantum operations plays important roles in the design of classical data hiding protocols~\cite{DuanFengYing2009} and the study of quantum reading capacity~\cite{das2017quantum}.

The discrimination protocol is a step-by-step procedure consisting of (the unknown) operation evaluations, along with quantum state preparations, additional quantum operations and measurements. The goal is to output the identity of the given operations, based on the measurement results. Comparing to the discrimination of quantum states, the discrimination of quantum operations admits more freedoms. To see this, we note that quantum operations are reusable, which enables quantum entanglement to be capitalized in the discrimination protocols. In addition, \emph{ancillary systems} are generally necessary for the optimal discrimination of two quantum operations. The perfect distinguishability of unitary operations~\cite{acin2001statistical} and quantum measurement apparatus~\cite{PhysRevLett.96.200401}, relies crucially on these aspects.

On the other hand, quantum operations can be used in many fundamental different ways, such as in parallel or in sequential. A parallel (discrimination) scheme enables the unknown quantum operation to be performed in parallel, which can be viewed as a direct generalization of the quantum state discrimination with multiple i.i.d. copies. A sequential scheme performs the unknown quantum operation step by step, while realizable extra quantum operations might be utilized to modify the intermediate states. Note that there exist quantum operations which cannot be distinguished using parallel schemes, but can be done by sequential schemes~\cite{PhysRevA.81.032339,duan2016parallel}. These two fundamental discrimination schemes turn out to be crucial in the study of the perfect distinguishability of quantum operations. Duan, Feng and Ying~\cite{DuanFengYing2009} concluded a sufficient and necessary condition to determine whether two quantum operations can be perfectly distinguished. In particular, for those perfectly distinguishable quantum operations, the discrimination protocol consists a finite number of uses of the unknown operations, and the application of extra quantum operations before performing measurements for the identifications.

When consider a restricted but important family of quantum operations -- the unitary gates (operations), the perfect discrimination among them is insensitive to the choice of strategies: Any two different unitary operations can be distinguished perfectly, by either applying the unknown one finite times in parallel~\cite{acin2001statistical}, or in sequential~\cite{duan2007entanglement}. Thus, there exists an interesting trade-off between the spatial resources (entanglement or circuits) and the temporal resources (running steps or discriminating times) in the discrimination of unitary operations~\cite{duan2007entanglement}. In principal, the main obstacle of performing parallel schemes is the difficulty of preparing pure multipartite entangled states. Performing sequential schemes can overcome this difficulty, while the long discriminating time may cause the \emph{decoherence}.

On experiment aspects, several pioneering experiments based on the non-universal devices have been devoted to related schemes. Liu and Hong~\cite{jian2008experimental} demonstrated the experiment on the sequential scheme  using Ti:Sapphire mode-locked laser. They reached successful probabilities around 99.5\% and 99.6\% respectively on two fixed examples. Zhang et al.~\cite{zhang2008linear} also used the laser performing the sequential protocol and reached the successful probabilities above 98\%. Laing, Rudolph and O'Brien~\cite{laing2009experimental} conducted the unitary quantum process discrimination (QPD) on photons without entanglement having a certainty around 99\% and the entanglement-assisted unitary QPD exceeding 97\% certainty.  

Although large-scale universal quantum computer may still be far off, we are approaching this so-called Noisy Intermediate Scale (NISQ) era of Quantum computing~\cite{preskill2018quantum}. In particular, IBM Corporation has started to provide quantum cloud service, called IBM Q. 
IBM Q enables us to perform high fidelity quantum gate operations and measurements on superconducting transmon qubits. In this paper, we implement both the parallel and sequential discrimination schemes to distinguish two qubit unitary gates, the $\frac{2}{3}\pi$-phase shift gate $R_{\frac{2}{3}\pi}=\begin{bmatrix}1&0\\0& e^{\frac{2}{3}i\pi}\end{bmatrix}$ and the identity gate $V=I=\begin{bmatrix}1&0\\0&1\end{bmatrix}$ on the $5$-qubit quantum processor (\emph{ibmqx4}). Note that $R_{\frac{2}{3}\pi}$ can be easily constructed using $QISKit$~\cite{cross2017open}. Moreover, we use the quantum programming platform $QSI$~\cite{liu2017q} to generate the discrimination schemes, determine the parameters of programs and translate to the quantum assembly language (QASM), which can be uploaded and performed on \emph{ibmqx4} via IBM Q cloud service.

In the following, we first present the parallel and sequential schemes to distinguish $R_{\frac{2}{3}\pi}$ and $I$, including the way to prepare the input states and perform measurements. Then, we exhibit the discrimination experiments performed on \emph{ibmqx4}~\cite{ibmqx2}, and analyze the (measurement) results. In the end, we discuss the advantages and disadvantages of parallel and sequential schemes, and propose some future directions.


\section{Description of the Experiments} 
\subsection{The discrimination schemes}
\paragraph{The Parallel Schemes}
As described in~\cite{acin2001statistical,duan2016parallel}, to distinguish two unitary gates, $R_{\frac{2}{3}\pi}$ and $V$, one may prepare an $N$-partite quantum states $\ket{\Psi}$ as the input for some positive integer $N$, such that $U^{\otimes N}\ket{\Psi}\perp V^{\otimes N}\ket{\Psi}$. To identify the unknown unitary operation, we perform the measurement $\{M_0=U^{\otimes N}\ket{\Psi}\!\bra{\Psi}(U^{\otimes N})^\dagger, M_1=V^{\otimes N}\ket{\Psi}\!\bra{\Psi}(V^{\otimes N})^\dagger\}$ if global operations are possible; otherwise we can implement the local discrimination protocol, introduced in~\cite{walgate2000local}. 
The outcome being $0$ corresponds to the unknown operation being $R_{\frac{2}{3}\pi}$; the outcome being $1$ corresponds to the unknown operation being $V$.

In our setting, we choose $N=2$ and the input state as
\begin{equation}\label{eq: parallel input}
\ket{\Psi}=(\frac{1}{\sqrt{3}}\ket{0}+\frac{1}{\sqrt{6}}\ket{1})\otimes\ket{0}+(-\frac{1}{\sqrt{6}}\ket{0}+\frac{1}{\sqrt{3}}\ket{1})\otimes\ket{1}\,.
\end{equation} 
It is easy to verify that 
$$R_{\frac{2}{3}\pi}^{\otimes 2}\ket{\Psi}=(\frac{1}{\sqrt{3}}\ket{0}+\frac{e^{\frac{2}{3}i\pi}}{\sqrt{6}}\ket{1})\otimes\ket{0}+(-\frac{e^{\frac{2}{3}i\pi}}{\sqrt{6}}\ket{0}+\frac{e^{\frac{4}{3}i\pi}}{\sqrt{3}}\ket{1})\otimes\ket{1}\,,$$
and $\bra{\Psi}R_{\frac{2}{3}\pi}^{\otimes 2}\ket{\Psi}=0$. 

\paragraph{The Sequential Schemes}
As described in~\cite{duan2007entanglement}, arbitrary two unitary operations, $R_{\frac{2}{3}\pi}$ and $V$, can be distinguished without entanglement, albeit additional unitary operations are required. Explicitly, we prepare $\ket{\Phi}$ as the input state, as well as a finite number of auxiliary unitary gates $X_1,\dots,X_{N-1}$. These auxiliary unitary gates will be applied to ensure that $UX_1U\cdots UX_{N-1}U\ket{\Phi}\perp VX_1V\cdots VX_{N-1}V\ket{\Phi}$. 

In our setting, only $1$ auxiliary unitary gate is required, which is the \emph{rotation matrix} $\begin{bmatrix}\cos \alpha&-\sin \alpha\\\sin \alpha&\cos \alpha\end{bmatrix}$ with $\alpha=\arctan(1/\sqrt{2})$. Explicitly,
$$X=\begin{bmatrix}\frac{\sqrt{2}}{\sqrt{3}}&-\frac{1}{\sqrt{3}}\\\frac{1}{\sqrt{3}}&\frac{\sqrt{2}}{\sqrt{3}}\end{bmatrix}\,.$$
Moreover, we choose the input as
\begin{equation}\label{eq: sequential input}
\ket{\Phi}:=\frac{1}{\sqrt{2}}(\ket{\varphi_0}+\ket{\varphi_1})\,,
\end{equation}
where $\ket{\varphi_0}$ and $\ket{\varphi_1}$ are the eigenvectors of 
\begin{equation*}
\begin{split}
X^\dagger UXU&=
\begin{bmatrix}\frac{1}{2}+\frac{\sqrt{3}i}{6}&-\frac{\sqrt{6}i}{3}\\-\frac{\sqrt{2}}{2}+\frac{\sqrt{6}i}{6}&-\frac{1}{2}-\frac{\sqrt{3}i}{6}\end{bmatrix}\,.
\end{split}
\end{equation*}
Eventually, we perform the measurement $\{M_0=UXU\ket{\Phi}\!\bra{\Phi}U^\dagger X^\dagger U^\dagger,M_1=X\ket{\Phi}\!\bra{\Phi}X^\dagger\}$. Resulting $0$ implies the unknown operation is $R_{\frac{2}{3}\pi}$, while resulting $1$ implies the unknown operation is $I$.


\subsection{Implementation Details}

\begin{figure}
\begin{subfigure}[b]{0.5\textwidth}
\includegraphics[width=\textwidth]{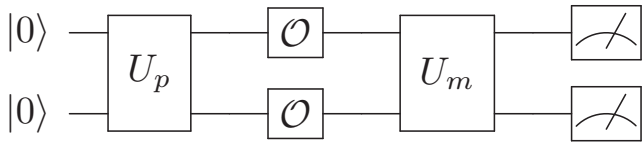}
\caption{The parallel scheme to distinguish the unknown operation $\mathcal{O}\in\{R_{\frac{2}{3}\pi},I\}$, where $U_p$ and $U_m$ indicate the state preparation and measurement circuits.}
\label{fig:parallel scheme}
\end{subfigure}\\[1ex]
\begin{subfigure}[b]{0.5\textwidth}
\centering
\includegraphics[width=\textwidth]{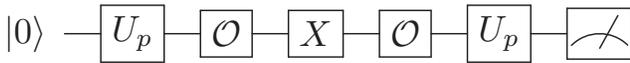}
\caption{The sequential scheme to distinguish the unknown operation $\mathcal{O}\in\{R_{\frac{2}{3}\pi},I\}$, where $U_p$ and $U_m$ indicate the state preparation and measurement circuits.}
\label{fig:sequential scheme}
\end{subfigure}
\caption{Parallel and sequential discrimination schemes}
\end{figure}
The parallel and sequential discrimination schemes are presented in FIG.~\ref{fig:parallel scheme} and FIG.~\ref{fig:sequential scheme}, respectively. Note that the unitary gate $R_{\frac{2}{3}\pi}=\begin{bmatrix}1&0\\0& e^{\frac{2}{3}i\pi}\end{bmatrix}$ can be generated by $QISKit$~\cite{cross2017open}. In fact, $QISKit$ can be used to implement all qubit unitary gates, parameterized as
\begin{equation*}
\begin{split}
U(\theta,\phi,\lambda)&:=\begin{bmatrix}
e^{-i(\phi+\lambda)/2}\cos(\theta/2)&-e^{-i(\phi-\lambda)/2}\sin(\theta/2)\\
e^{i(\phi-\lambda)/2}\sin(\theta/2)&e^{i(\phi+\lambda)/2}\cos(\theta/2)
\end{bmatrix}
\end{split}
\end{equation*}
on the quantum processor with gate fidelity around $99.9\%$. Note that IBM's quantum processor only supports that each qubit is initialized to $\ket{0}$, and measure each qubit with respect to the computational basis $\{\proj{0},\proj{1}\}$. Thus, we need to generate the input state preparation circuits and rotate the measurement to computational basis. In the sequential scheme (FIG.~\ref{fig:sequential scheme}), $U_p=U(1.1503,6.4850,2.2555)$ and $U_m=U(0.7854,6.0214,6.1913)$. Implementing the circuit in FIG.~\ref{fig:sequential scheme} and measuring the output state, we assert that $\mathcal{O}$ is $R_{\frac{2}{3}\pi}$ if the (measurement) output is $0$; $\mathcal{O}$ is $I$ if the output is $1$.

\begin{figure}
\includegraphics[width=0.5\textwidth]{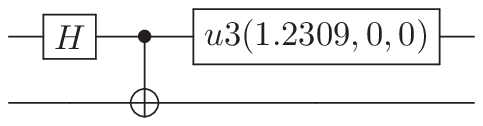}
\caption{The quantum circuit ($U_p$) which generate $\ket{\Psi}$ from $\ket{0}\otimes\ket{0}$.}
\label{fig: parallel input}
\end{figure}

\begin{figure}
\includegraphics[width=0.5\textwidth]{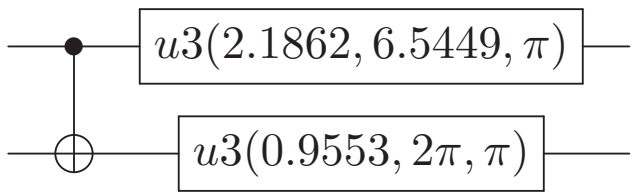}
\caption{The quantum circuit ($U_m$) which distinguish $U^{\otimes 2}\ket{\Psi}$ and $\ket{\Psi}$.}
\label{fig: parallel measurement}
\end{figure}

In the parallel scheme, to prepare the input state $\ket{\Psi}$, computed in Eq.~\ref{eq: parallel input}, we utilize the circuit presented in FIG.~\ref{fig: parallel input}. 
In the measurement step, we implement the local discrimination protocol for two multipartite states~\cite{walgate2000local}, as shown in FIG.~\ref{fig: parallel measurement}. Implementing such a circuit and measuring the output state, we say $\mathcal{O}$ is $R_{\frac{2}{3}\pi}$ if the output is $01$ or $10$; and $\mathcal{O}$ is $I$ if the output is $00$ or $11$.

\section{The experiments}
We perform the discrimination experiments on the IBM's quantum processor \emph{ibmqx4}, while generate the circuits by $QSI$ (the key code segments can be found in (\url{https://github.com/klinus9542/UnitaryDistIBMQ})).
To simulate the secret chosen procedure, we simply generate a uniformly random bit for choosing the identity of $I$ and $R_{\frac{2}{3}\pi}$, which can be accomplished in $QSI$ easily. Then we generate the discrimination protocols, as shown in FIG.~\ref{fig:parallel scheme} and FIG.~\ref{fig:sequential scheme} replacing the gate $\mathcal{O}$ by the chosen gate. $QSI$ converts the quantum circuit to the quantum assembly language, and execute the experiments on \emph{ibmqx4} through the application programming interface (API) of quantum cloud service provided by IBM. 
For each random bit, we execute the discrimination scheme on \emph{ibmqx4} for $1024$ times and gather the measurement results. 

Based on the theoretical calculations, the identity of the chosen unitary gates will be perfectly determined. For instance, when we apply parallel scheme (FIG.~\ref{fig:parallel scheme}) and $\mathcal{O}$ is chosen as $R_{\frac{2}{3}\pi}$, the measurement outputs should only contains $01$ and $10$, which appears with equally many times. However, current quantum technologies may not be able to achieve the theoretical performance. As mentioned before, the fidelity of single qubit gate is still not perfect, which causes unavoidable error. Another type of error arises from introducing the state preparation circuits and measurement circuits since the theoretical input states and the measurements contain \emph{irrational} parameters presented by float type in software, which cannot be created accurately. Last but not least, the measurement results need to be sorted, as some ``impossible'' results might appear: In principal, the statistical results can be $xy000$ when using the 5-qubit \emph{ibmqx4} chip. However, in fact, the outputs can be arbitrary $5$-bit strings as there might be errors between used qubits and unused qubits. For these, we ignore the unused qubits and sort the final results.  

\begin{figure}[!ht]
	\centering
		\begin{subfigure}[b]{0.5\textwidth}
		\centering
		\includegraphics[width=\textwidth]{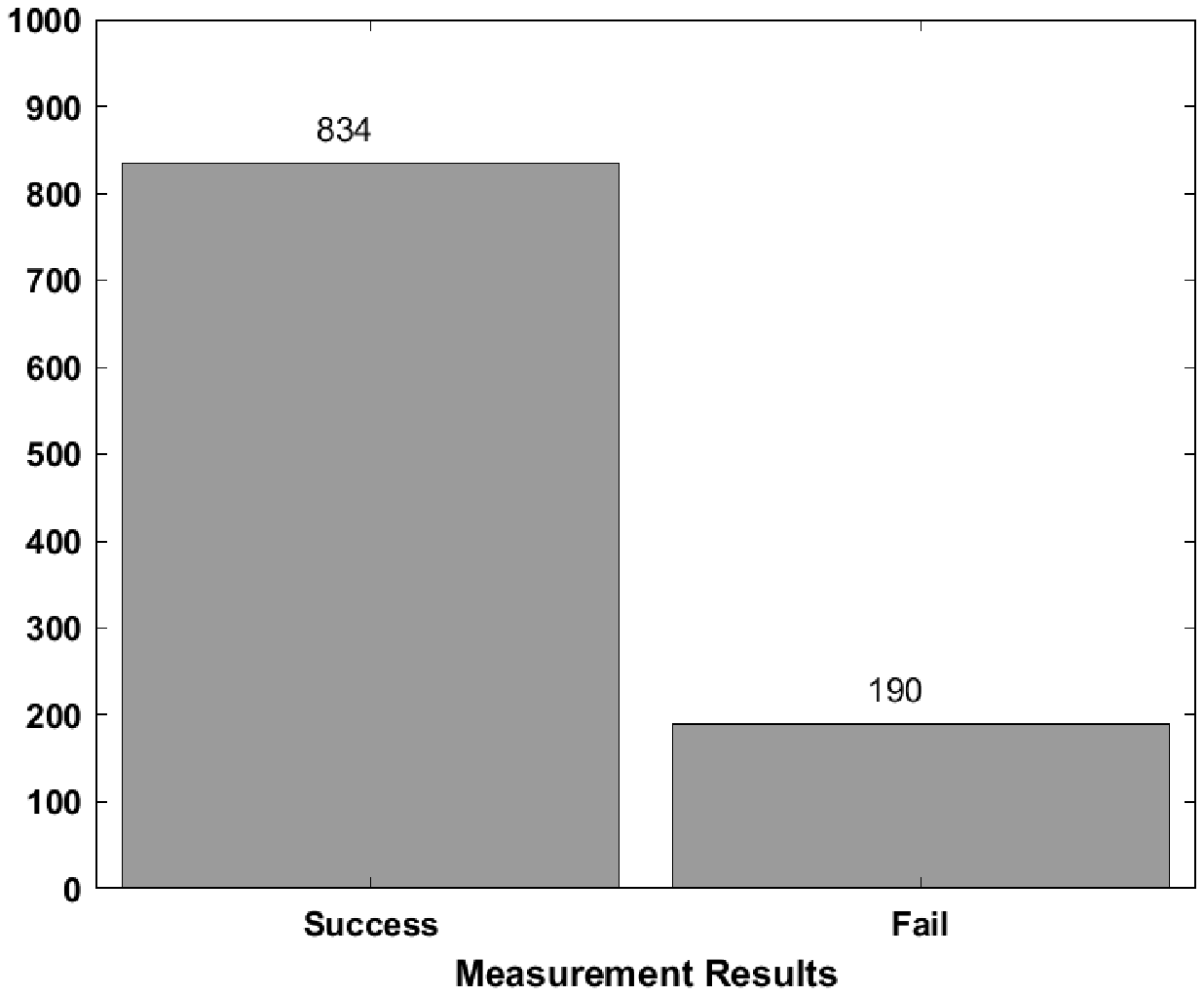}
		\caption{Perform the circuit in FIG.~\ref{fig:parallel scheme} by replacing $\mathcal{O}$ by $R_{\frac{2}{3}\pi}$ for 1024 times. After sorting the outputs, $834$ rounds output either $01$ or $10$ (indicates $\mathcal{O}$ is $R_{\frac{2}{3}\pi}$), and $190$ rounds output either $00$ or $11$ or other results (indicate $\mathcal{O}$ is not $R_{\frac{2}{3}\pi}$).}
		\label{fig:ParOneSettingU}
	\end{subfigure}
	\begin{subfigure}[b]{0.5\textwidth}
		\centering
		\includegraphics[width=\textwidth]{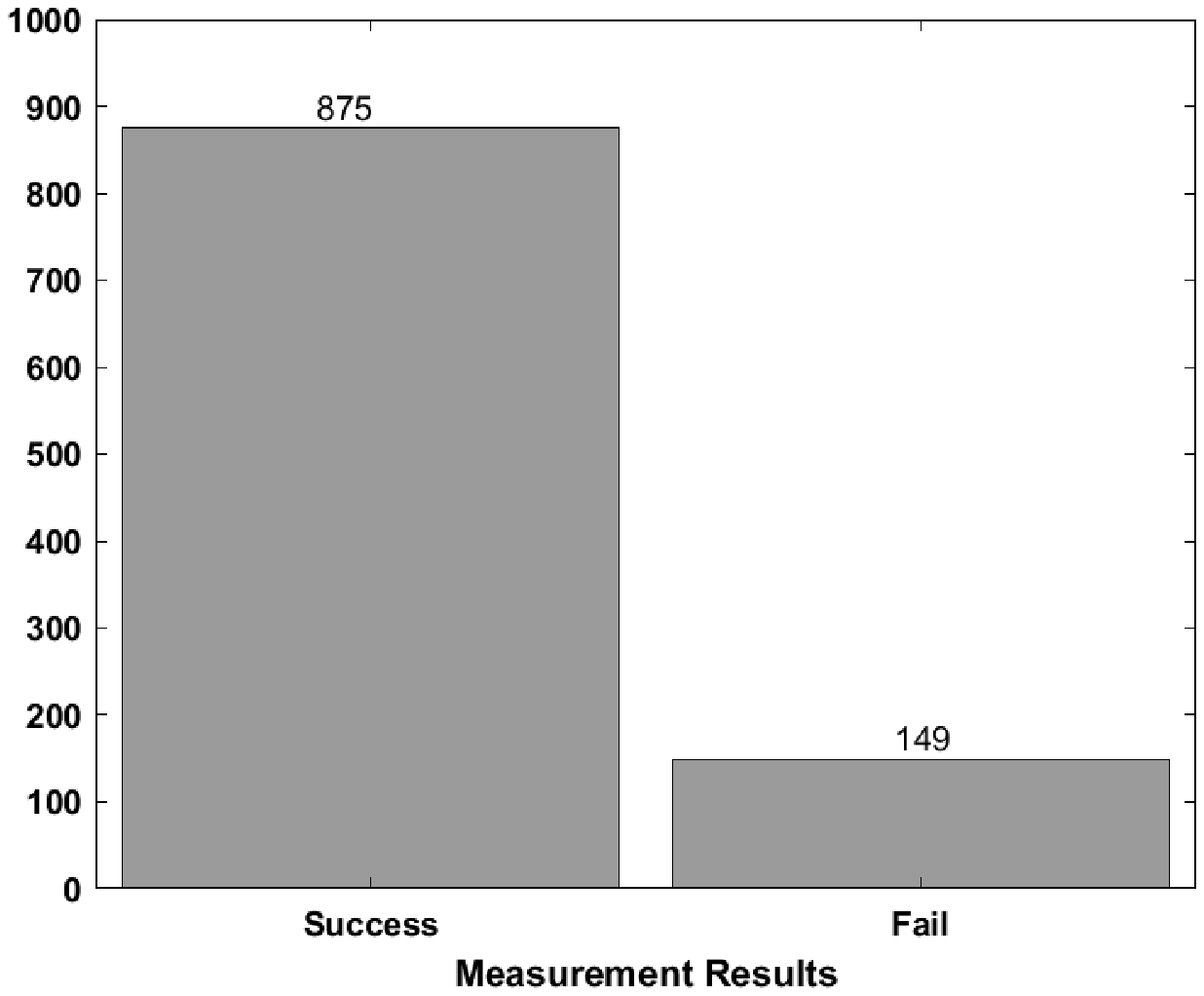}
		\caption{Perform the circuit in FIG.~\ref{fig:parallel scheme} by replacing $\mathcal{O}$ by $I$ for 1024 times. After sorting the outputs, $875$ rounds output either $00$ or $11$ (indicate $\mathcal{O}$ is $I$), and $149$ rounds output either $01$ or $10$ or other results (indicate $\mathcal{O}$ is not $I$). }
		\label{fig:ParOneSettingI}
	\end{subfigure}	
	\caption{Statistical results in the parallel discrimination experiments.}
		\label{fig:parSetting}
\end{figure}
\begin{figure}[!ht]        
	\centering   
	\begin{subfigure}[b]{0.5\textwidth}
		\centering 
		\includegraphics[width=\textwidth]{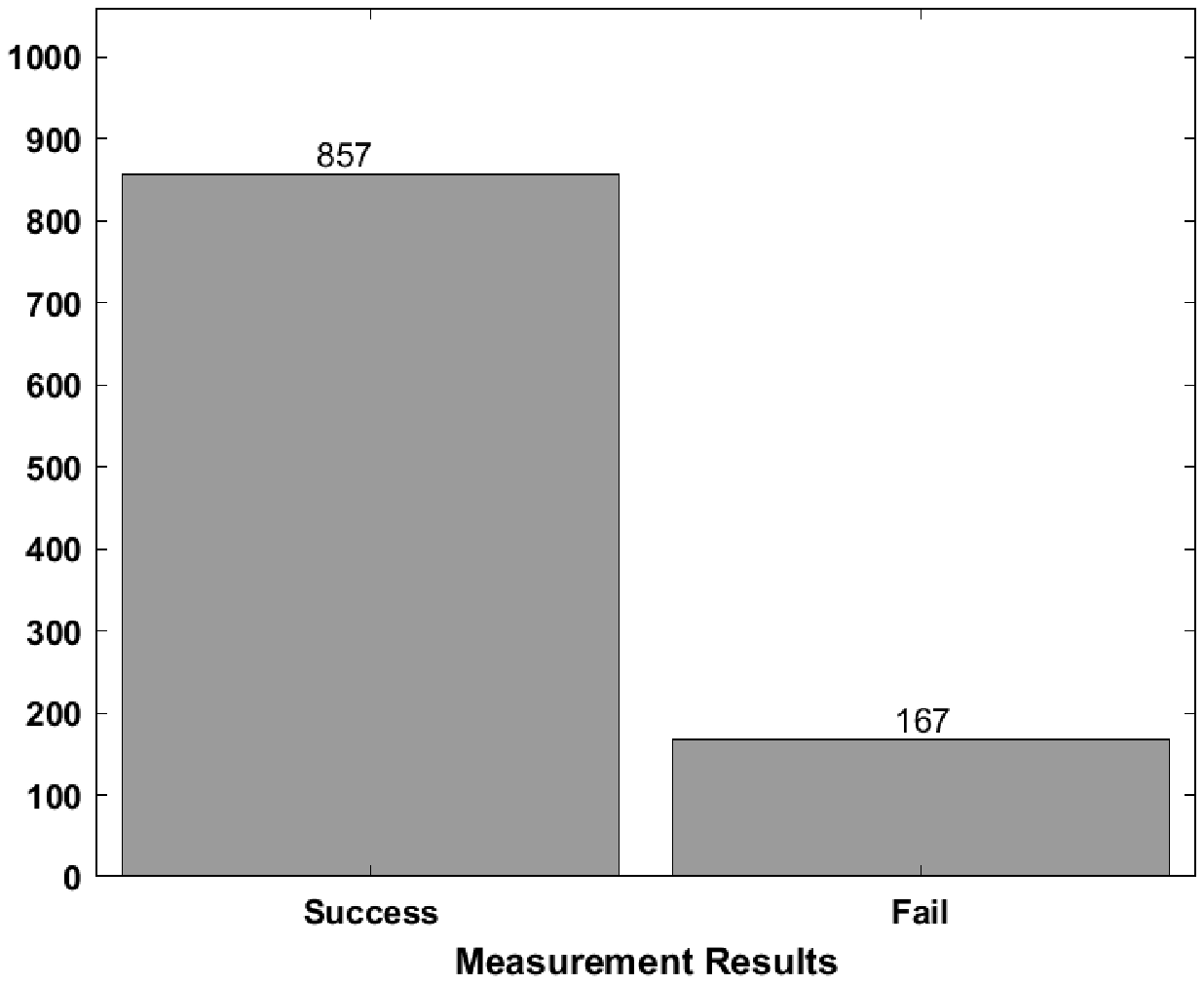}
		\caption{Perform the circuit in FIG.~\ref{fig:sequential scheme} by replacing $\mathcal{O}$ by $R_{\frac{2}{3}\pi}$ for $1024$ times. After sorting the outputs, $857$ rounds output $0$ (indicate $\mathcal{O}$ is $R_{\frac{2}{3}\pi}$), and $167$ rounds output either $1$ or other results (indicate $\mathcal{O}$ is not $R_{\frac{2}{3}\pi}$). }
		\label{fig:SeqOneSettingU}
	\end{subfigure}\\[1ex]
	\begin{subfigure}[b]{0.5\textwidth}
		\centering
		\includegraphics[width=\textwidth]{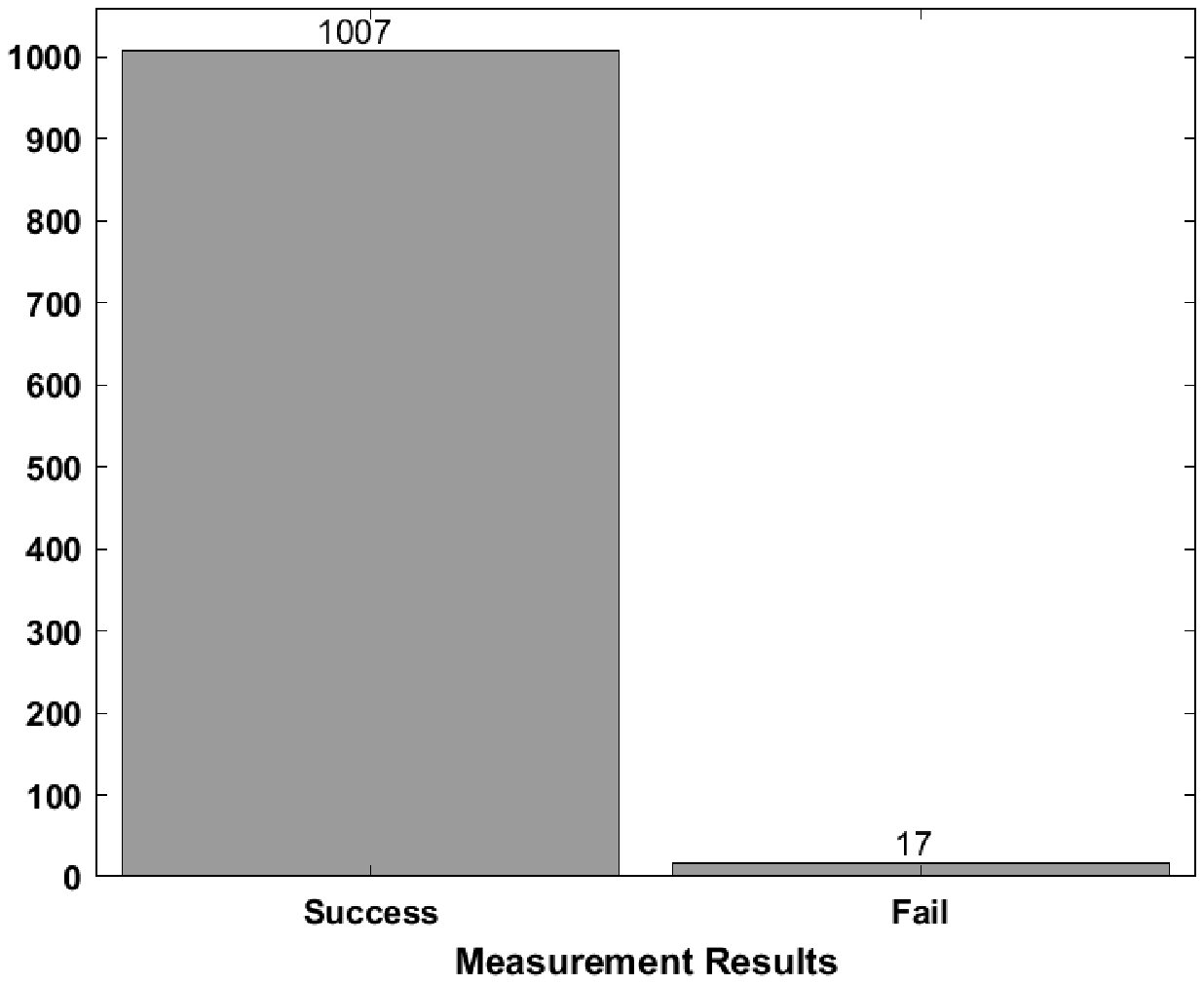}
		\caption{Perform the circuit in FIG.~\ref{fig:sequential scheme} by replacing $\mathcal{O}$ by $I$ for $1024$ times. After sorting the outputs, $1007$ rounds output $1$ (indicate $\mathcal{O}$ is $I$), and $17$ rounds output either $1$ or other results (indicate $\mathcal{O}$ is not $I$). }
		\label{fig:SeqOneSettingI}
	\end{subfigure}
	\caption{Statistical results in the sequential discrimination experiments.}
	\label{fig: seqSetting}
\end{figure}

\begin{figure}[!ht]  
	\includegraphics[width=0.5\textwidth]{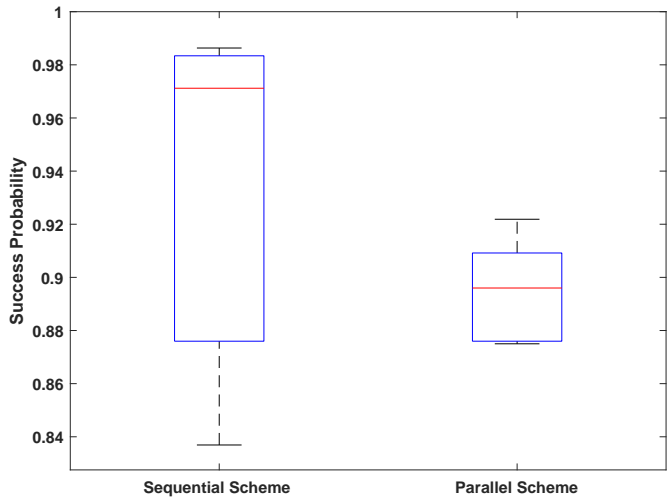}  
	\caption{The discrimination success probability distributions for both sequential and parallel discrimination. For each round in each scheme, $R_{\frac{2}{3}\pi}$ and $I$ are chosen depending on a random coin-flip result. For each scheme, we execute the experiment for $10$ randomly chosen $\mathcal{O}$. In each box, the central mark indicates the median, and the top and the bottom indicate the $75\%$ and $25\%$ percentiles, respectively.}  
	\label{fig:SeqParMultiShots}  
\end{figure}

FIG.~\ref{fig:ParOneSettingU} and FIG.~\ref{fig:ParOneSettingI} stand for the statistical measurement results for parallel discrimination schemes, and FIG.~\ref{fig:SeqOneSettingU} and FIG.~\ref{fig:SeqOneSettingI} stand for the statistical measurement results for parallel discrimination schemes. FIG.~\ref{fig:SeqParMultiShots} illustrates the box-plot of success probabilities on parallel and sequential schemes, where we perform each scheme $10$ times with randomly chosen $\mathcal{O}$, each of which includes $1024$ repeating experiments. The choices of $\mathcal{O}$ depend on the value of a random bit, generated on classical computers. It can be observed that both the worst ($85.83\%$) and the best ($98.63\%$) success probabilities come from the sequential discrimination experiments. In particular, the best success probability is achieved when $\mathcal{O}$ is replaced by $I$. Thus, the discrimination scheme (FIG.~\ref{fig:sequential scheme}) contains only three qubit gates. On the other hand, the worst success probability is achieved when $\mathcal{O}$ is replaced by $R_{\frac{2}{3}\pi}$, where $5$ (rather complicated)  gates need to be executed, which might increase the error. For the parallel scheme, the success probabilities are ranging from $88\%$ to $92\%$, with not very significant differences (stand deviation of parallel scheme is $\sigma=0.017$, compared with stand deviation of sequential scheme is $\sigma=0.061$).

\section{Conclusion and Discussion}\label{sec: dis}
In this paper, we distinguish unitary gates by parallel scheme and sequential scheme on the IBM's quantum processor \emph{ibmqx4}. Both two schemes are proposed to achieve the perfect discrimination theoretically. In our experiments, we report that both two schemes can distinguish the qubit unitary gates $R_{\frac{2}{3}\pi}$ and $I$ with success probability over $85\%$, under the condition of superconducting universal quantum computer. In addition, we utilize $QSI$ modules to perform $10$ random experiments for parallel scheme and sequential scheme, each of which chooses $R_{\frac{2}{3}\pi}$ and $I$ uniformly at random. FIG.~\ref{fig:SeqParMultiShots} suggests both two schemes can distinguish the randomly chosen unitary gates with high probabilities. Moreover, we infer that using the sequential scheme may achieve higher success probabilities than the parallel scheme, while the success probabilities using parallel scheme are more robust than using sequential schemes. In particular, when the set of known unitary gates are with rather simple structures, such as the identity gate or Hadamard gate, the sequential scheme admits more advantages in the discriminations. We assert that this is due to the fact that the coherence and fidelity of two-qubits gates are still not ideal in IBM quantum processors. On the other hand, using parallel discrimination scheme is more robust: it may not achieve a $90\%$ success probability, while the success probabilities do not differ too much. We left implementing the discrimination of general quantum operations as a further direction.




\section*{Acknowledgments}
The authors were grateful to the use of the IBM Q experience, and acknowledge IBM Q community for their helpful discussions. The views expressed are those of the authors and do not reflect the official policy or position of IBM or the IBM Q experience team. SL is supported by the National Natural Science Foundation of China (Grant No.61672007). YL is supported by ERC Consolidator Grant 615307-QPROGRESS. 

\bibliography{ref}
\bibliographystyle{apsrev4-1}
\end{document}